\documentclass[a4paper, 10 pt, conference]{ieeeconf}
\IEEEoverridecommandlockouts
\usepackage{tikz}
\usepackage{bm}
\usepackage{pgfplots}
\usepackage{booktabs}
\usepackage{adjustbox}
\usepackage{algorithm,algorithmic}

\pgfplotsset{compat=1.3}

\usepackage{amsmath,mathtools,{amssymb}}
\usepackage{blkarray}
\DeclareMathOperator{\diag}{diag}
\DeclareMathOperator{\RGA}{RGA}
\DeclareMathOperator{\sgn}{sgn}

\usepackage{nicematrix}
\NiceMatrixOptions{
code-for-first-row = \color{blue},
code-for-last-row = \color{blue},
code-for-first-col = \color{blue},
code-for-last-col = \color{blue}
}

\newcommand{\markr}[0]{\textcolor{red}{$\bigstar$}}
\newcommand{\markb}[0]{\textcolor{blue}{$\blacksquare$}}
\newcommand{\rt}[1]{\textcolor{red}{($#1\%$)}}

\newcommand{\pr}[0]{\phantom{\markr}}
\newcommand{\pb}[0]{\phantom{\markb}}

\title{\LARGE \bf
A Simple Discretization Scheme for Gain Matrix Conditioning
}

\author{Daniel L. O'Connor$^{1}$, Lim C. Siang$^{2,3}$ and Shams Elnawawi$^{2}$% <-this % stops a space
\thanks{$^{1}$Control Consulting Inc., Great Falls, MT, United States}
\thanks{$^{2}$Department of Process Control Engineering, Burnaby Refinery, BC, Canada}
\thanks{$^{3}$College of Computing, Georgia Institute of Technology, Atlanta, GA, United States}
}

\begin{document}

\maketitle
\thispagestyle{empty}
\pagestyle{empty}

\begin{abstract}

% Dynamic models used in industrial model predictive controllers (MPCs) are commonly generated from regression-based system identification methods. The usage of these regression-based methods to generate the steady-state gain matrix will maximize the gain matrix's degrees of freedom based on empirical plant data, and the result is a 

In industrial model predictive controllers (MPCs), models generated from regression-based system identification methods typically contain small or even physically non-existent degrees of freedom.  Control issues can arise when the steady-state optimizer uses these small degrees of freedom to calculate targets for plant operation due to matrix ill-conditioning. Mathematical techniques like Relative Gain Array (RGA) and Singular Value Decomposition (SVD) are helpful for analyzing controller gain interactions and identifying conditioning issues, which can be corrected relatively easily in small models.  However, these techniques are difficult and tedious to apply for larger, more complex models. This paper describes a novel, non-iterative, RGA-based, binning technique for discretizing the gain matrix and quickly solving 2$\times$2 conditioning issues for any model size, while guaranteeing gain adjustments below a certain threshold. Higher order interactions are also discussed.
\end{abstract}


\section{INTRODUCTION}

Industrial model predictive controllers (MPCs) are commonly implemented with an integrated steady-state, Linear Program (LP) optimizer to generate targets that drive the process to an economic optimum \cite{maciejowskiPredictiveControlConstraints2002}. In industrial MPC projects, one of the most difficult and time-consuming steps is understanding the interaction of the steady-state model gains with the optimizer costs in the LP solution. Part of this consideration is respecting the material and energy balance of the process, and determining how much ``detail" in terms of smaller gains should be built into the model.

In practice, designers must make structural decisions in the gain matrix to reconcile the model with the true process. This step relies on the designer's process knowledge and experience, along with the usage of process data to justify any curve or gain manipulation of the model. Adjusting the gain matrix is often a balancing act between maintaining model accuracy and usability, and arbitrarily deleting or scaling gains should be avoided. A case study from Mitsubishi demonstrated how the erroneous removal of a gain element, thought to be insignificant based on engineering judgment, resulted in controller instabilities \cite{ishikawaPracticalMethodRemoving1997}.

% The gains and relative gains will affect the APC controller dynamic move calculations. If the gains are calling for a large overall move in an MV from its current value to the \textit{steady state target} calculated by the LP, the current move is also likely to be large. If they are wrong, it’s difficult to reduce the move size by tuning the controller. A good example of this are non-linear CVs such as valve output constraints or distillation column dPs.  Here, the gains change depending on the value of the CV.  For these types of problems, linearizing transformations are applied. These are structural gain problems that are fixed by linearization rather than gain manipulation.  

Fundamentally, there are 3 types of MV-CV (Manipulated Variable-Controlled Variable) interactions:

\begin{itemize}
    \item \textbf{Non-collinearities:} MV-CV interactions that are easy to independently control
    \item \textbf{Collinearities:} MV-CV interactions that cannot be independently controlled, because the gains are collinear
    \item \textbf{Near-collinearities:} MV-CV interactions that are difficult to independently control
\end{itemize}

The non-collinear MV-CV interactions are typically trivial to model and tune, and do not warrant in-depth discussions for the purposes of this paper.

\begin{figure}
    \centering
    \includegraphics[width=0.90\linewidth]{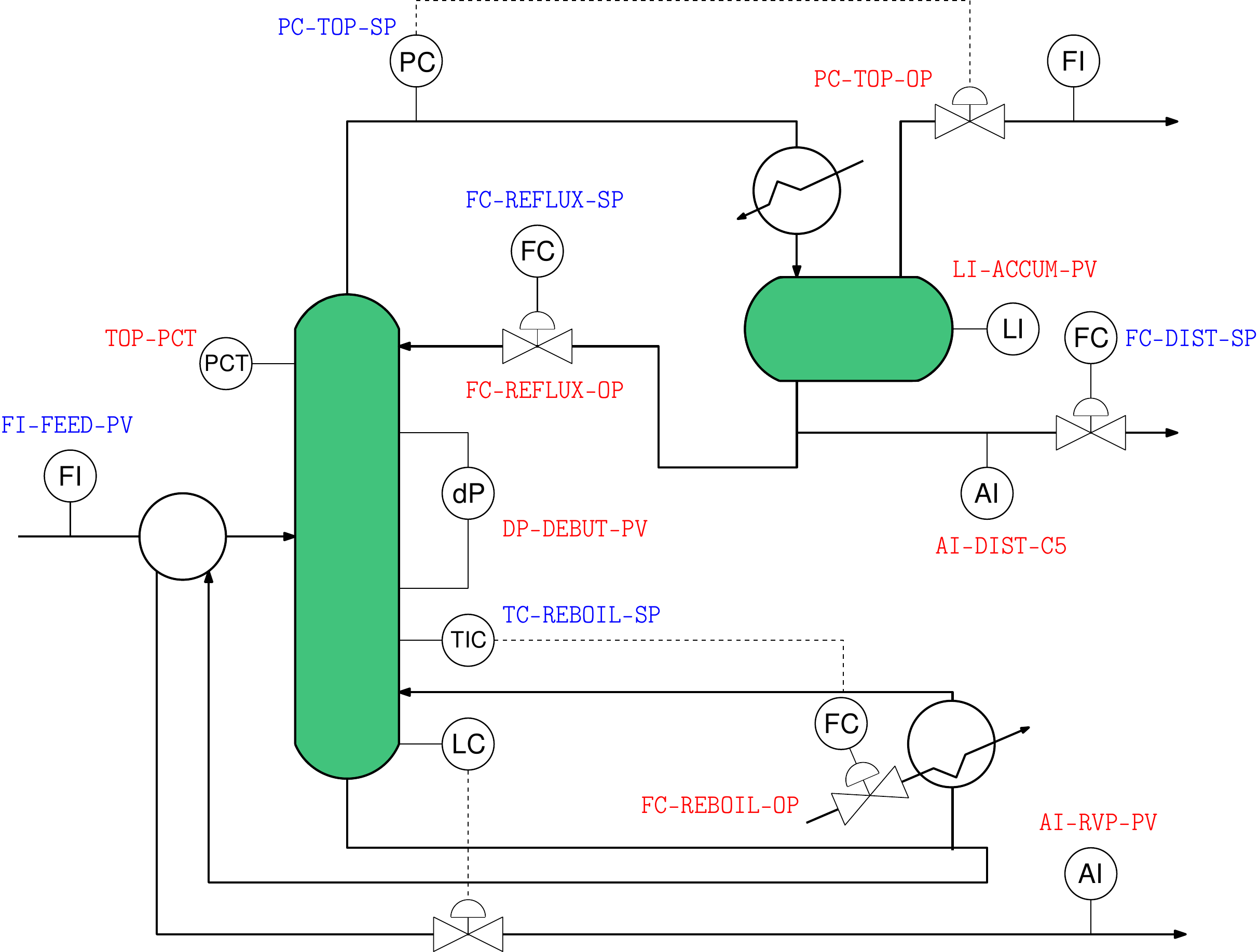}
    \caption{Process flow diagram (PFD) for a debutanizer column, which separates butane and light-end components from a heavier (C5+) hydrocarbon mixture. MVs are highlighted in blue, while CVs are red.}
    % The column feed is preheated using bottoms product. Overhead product is condensed and separated in a reflux drum, where the reflux flow is used to control overhead product composition. Similarly, the bottoms composition is controlled via the reboiler heat duty. 
    \label{fig:debutanizer}
\end{figure}

The collinear MV-CV interactions can often be identified based on process data and \textit{a priori} process knowledge. These interactions may not be exactly collinear in the raw model, but can be corrected by forcing the gain ratios of the collinear MV-CV pairs to be numerically equal. Doing so ensures that the LP will not try to use control handles that are nonexistent.

% step, it is highly unlikely that the raw gains from system identification will be precisely collinear. 

The near-collinear MV-CV pairs can result in problematic interactions due to numerically ill-conditioned matrices, which, if active, cause undesirable large MV movements to control the near-collinear CVs independently at their constraint limits. In industry, it is common for control practitioners to adjust the steady-state gains manually \cite{ishikawaPracticalMethodRemoving1997} to deal with ill-conditioning issues. When a gain is changed, its interaction with all of the other gains also changes. Consequently, fixing one gain conditioning issue may create other issues in other parts of the matrix. For further details, Waller and Waller (1995) \cite{wallerDefiningDirectionalityUse1995a} provides an excellent review of process directionality literature.

An ExxonMobil patent \cite{hallMethodModelGain2008} describes a mathematical procedure for adjusting gain matrices. Some commercial applications try to automatically correct the ill-conditioning, such as Honeywell's RMPCT which uses Singular Value Thresholding (SVT) \cite{qinSurveyIndustrialModel2003}, but automatic techniques can fail because the raw gains can be far enough from the collinearity threshold that they are not fixed. On the other hand, making the SVT threshold large enough to fix these gains could incorrectly remove small but critical interactions/degrees of freedom. Semi-automated techniques, such as those described in a patent by Aspentech \cite{zhengMethodsArticlesDetecting2014}, may not always produce satisfactory results that are consistent with the process, and may warrant further manual adjustments.

In this paper, we present a novel, non-iterative, binning approach to gain conditioning that significantly reduces the difficulties associated with gain matrix adjustments faced by control practitioners.

\section{REVIEW OF TECHNIQUES}

Two primary tools for analyzing model conditioning are Singular Value Decomposition (SVD), and the Relative Gain Array (RGA).

\subsection{Singular Value Decomposition}
Consider a $2\times2$ steady-state gain matrix, $G$,

% \begin{equation}
% \label{eqn:svd}
% G = U \Sigma V^\mathsf{T}
% \end{equation}

\begin{equation}
\label{eqn:ssgain-general}
G =
    \begin{bNiceMatrix}[first-row, last-col=3]
        {\footnotesize{\textrm{MV1}}} &  {\footnotesize{\textrm{MV2}}} & \\
        G_{11} & G_{12} & \footnotesize{\textrm{CV1}}  \\
        G_{21} & G_{22} & \footnotesize{\textrm{CV2}}
    \end{bNiceMatrix}
\end{equation}

SVD decomposes $G$ into three matrices with $G = U \Sigma V^\mathsf{T}$, describing its most responsive and least responsive directions in terms of both MV moves and CV responses, where
\begin{equation*}
    \Sigma = 
    \begin{bNiceMatrix}
        \overline{\sigma} & 0  \\
        0 & \underline{\sigma}
    \end{bNiceMatrix} \quad
    U = 
    \begin{bNiceMatrix}[first-row]
        {\overline{\bm{u}}} &  {\underline{\bm{u}}} \\
        u_{11} & u_{12} \\
        u_{21} & u_{22}
    \end{bNiceMatrix}
    \quad
    V^\mathsf{T} = 
    \begin{bNiceMatrix}[last-col]
        v_{11}^\mathsf{T} & v_{12}^\mathsf{T} & \overline{\bm{v}}^\mathsf{T} \\
        v_{21}^\mathsf{T} & v_{22}^\mathsf{T} & \underline{\bm{v}}^\mathsf{T}
    \end{bNiceMatrix}
\end{equation*} % https://tex.stackexchange.com/questions/59517/label-rows-of-a-matrix-by-characters

\begin{figure}
    \centering
    \includegraphics[width=0.48\textwidth]{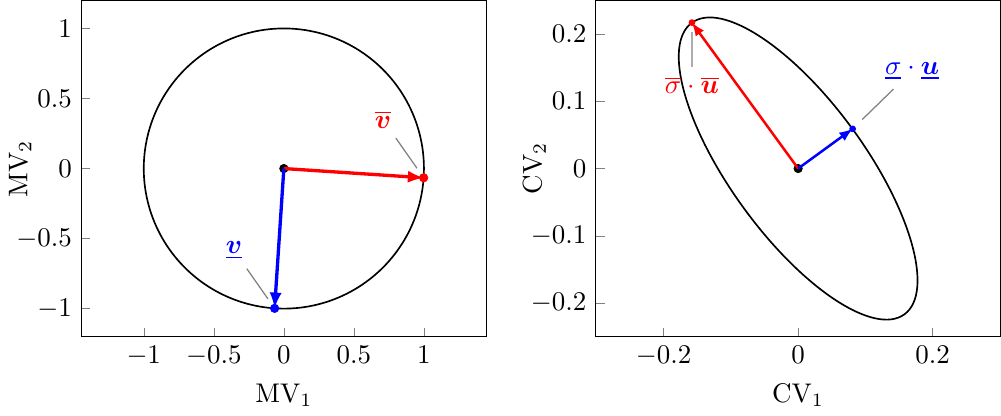}
    \caption{Graphical interpretation of SVD. The left subfigure represents a unit circle of MV input moves, and the right subfigure represents the CV responses. SVD breaks down the CV responses in terms of a `strong' (red) and `weak' (blue) direction illustrated by the CV ellipse. As the weak direction becomes weaker, $\gamma$ increases and the system converges towards collinearity. A collinear matrix would collapse the CV ellipse into a line in the direction of $\overline{\bm{u}}$. }
    % $U$ is shown in the directions of the long and short sides of the ellipse on the right, where $\Sigma$ represents the relative lengths of these sides. 
    \label{fig:svd_unscaled}
\end{figure}

A physical interpretation of SVD in the context of multivariable control is described in detail by Moore (1986) \cite{mooreApplicationSingularValue1986}, as illustrated in Fig. \ref{fig:svd_unscaled}. CVs will be most responsive in the direction of $\overline{\bm{u}}$ when the MVs are moved in the direction of $\overline{\bm{v}}^\mathsf{T}$. The opposite is true for $\underline{\bm{u}}$ and $\underline{\bm{v}}^\mathsf{T}$. The singular values $(\overline{\sigma}, \underline{\sigma})$ represent the magnitude of each CV direction's controllability.

Singular values are proportional to system sensitivity \cite{mooreApplicationSingularValue1986}; high $\sigma$ values indicate that MV moves will need to be small to maintain accurate control, and vice versa. Singular values that are close to each other indicate that, generally, a controller will be able to control different CVs relatively independently. In contrast, systems with very different singular values will require larger MV moves for disproportionately smaller CV changes. This effect is encapsulated by the condition number, which is defined as the ratio of the largest to the smallest singular value:

% For SISO control, singular value is represented by the steady-state gain, and any gain can be accounted for in the design of the controller, although very large gains may result in loss of resolution in the controller output. However, the presence of interactions in MIMO control makes the singular values a more significant factor in maintaining control. 

\begin{equation}
\gamma = {\overline{\sigma}}/{\underline{\sigma}}
\end{equation}

% where the column vectors, $U = \left[ \overline{\bm{u}}, \underline{\bm{u}} \right]$ represents the two orthogonal CV output directions for the matrix, the diagonal matrix $\Sigma = \diag \left(\overline{\sigma}, \underline{\sigma} \right)$ are the singular values, which represent the magnitude the process drives in each CV direction, and the row vector $V^\mathsf{T} = \left[ \overline{\bm{v}}^\mathsf{T},\underline{\bm{v}}^\mathsf{T} \right]$ represents the orthogonal MV move sizes associated with the orthogonal CV directions (input directions).

% A gain matrix based on the debutanizer process in Figure \ref{fig:debutanizer} is shown in Eq. (\ref{eqn:ssgain}). We consider just the reflux flow rate and reboiler temperature control MVs, and the overhead C5’s and bottoms Reid Vapour Pressure (RVP) CVs, and illustrate how SVD helps us analyze the relative gains.

It is important to note that the SVD is scale-dependent. The singular values and condition numbers calculated are based on the engineering units of the MVs and CVs.

As an example, consider an unscaled gain matrix in Eq. (\ref{eqn:ssgain}) based on the debutanizer process in Figure \ref{fig:debutanizer}. \textrm{MV1} is the \texttt{TC-REBOILER-SP}, \textrm{MV2} is the \texttt{FC-REFLUX-SP}, \textrm{CV1} is the \texttt{AI-RVP-PV} in the bottoms and \textrm{CV2} is the \texttt{AI-DIST-C5} in the overhead. 

% We will use this example to illustrate how SVD helps us analyze the relative gains.

\begin{equation}
\label{eqn:ssgain}
G =
    \begin{bNiceMatrix}[first-row, last-col=3]
        {\footnotesize{\textrm{MV1}}} &  {\footnotesize{\textrm{MV2}}} & \\
        -0.1942 & -0.0029 & \footnotesize{\textrm{CV1}}  \\
        0.1843 & -0.0288 & \footnotesize{\textrm{CV2}}
    \end{bNiceMatrix}
\end{equation}

% \begin{equation}
% \label{eqn:ssgain}
% \renewcommand\arraystretch{1.5}
% \begin{blockarray}{rrr}
%                  & \rot{\texttt{TC-REBOILER}} &  \rot{\texttt{FC-REFLUX}} \\
%   \begin{block}{r[rr]}
%         \texttt{AI-BTMS-RVP}
%         & -0.1942 & -0.0029  \\
%         \texttt{AI-OVHD-C5} & 0.1843 & -0.0288 \\
%   \end{block}
% \end{blockarray}
% \end{equation}

The SVD results based on Eq. (\ref{eqn:ssgain}) are summarized below. 

\begin{equation*}
U =
\begin{bmatrix*}[r]
-0.7215 & 0.6925 \\
0.6925 & 0.7215
\end{bmatrix*}
\quad
V = 
\begin{bmatrix*}[r]
0.9978 & -0.0664 \\
-0.0664 & -0.9978
\end{bmatrix*}
\end{equation*}

\begin{equation*}
\Sigma =
\begin{bmatrix*}[c]
0.2683 & 0 \\
0 & 0.0229
\end{bmatrix*}
\end{equation*}

The condition number is calculated to be $\gamma = 11.74$. However, due to the scale-dependent properties of SVD, the results obtained with Eq. (\ref{eqn:ssgain}) may not be meaningful.  

% The graphical representation of these SVD operations is also shown in Fig. \ref{fig:svd_unscaled}, showing what each of the $U$, $V$ and $\Sigma$ matrices represents.

% Consider a unit circle of moves for the MV’s, and let’s plot the resulting CV pattern.  

% The method takes either the average or maximum move size for an MV made during the plant test, multiplies the CV gains associated with that MV by that move size, and then divides each CV by the maximum gain of each CV.

Researchers have proposed various scaling choices, including scaling by physical limits, relative importance of variables, or by other problem-specific scaling that minimizes the condition number to obtain $\gamma_{\min}$ \cite{wallerDefiningDirectionalityUse1995a}. In practice, a common and useful way to scale the matrix is \textit{typical move scaling}.
This method multiplies each gain by the maximum MV move size made during identification, then divides it by the maximum gain of that CV across all other MVs. The scaled matrix, $G'$ using typical move scaling is given by
\begin{equation}
% Notation used in Waller and Waller
% https://math.stackexchange.com/questions/3274790/notation-for-element-wise-multiplication-of-vector-and-matrix-columns
\label{eqn:svd_scaling}
G' = S_b G S_a
\end{equation}
where $S_a$ and $S_b$ are non-zero, diagonal scaling matrices such that
$
S_a = \diag\left(\Delta\textrm{MV}_1, \Delta\textrm{MV}_2\right)
$
and
$
S_b = \diag(\underset{i}{\max} (|GS_{a}|)_{1, i}, \underset{i}{\max} (|GS_{a}|)_{2, i})^{-1}
$.
In the first scaling step, the diagonal elements in $S_a$ scale each MV column in $G$ by its $\Delta\textrm{MV}$ move size. In the second scaling step, the diagonal elements in $S_b$ scale each CV row by dividing each row by that CV's maximum gain in the $GS_a$ matrix.

% Note that $GS_a$ is used because this directly calculates the CV responses in terms of the MV moves, rather than expressing $S_b$ in terms of CV responses.

Typical move scaling `normalizes' the gains by considering both MV move sizes and CV responses to all related MVs. Consequently, small scaled gains may indicate that the gain is small or insignificant, or the MV move size was too small for that CV, or that large MV moves were made with a high-gain CV, making other CV scaled gains look small.

% The smaller the gain, the less significant the CV response based on the move sizes made.  Typical move scaling doesn’t necessarily mean a small gain is insignificant, it could mean the MV move size was too small to generate a significant response in that CV, or really big moves were made in an MV with a large CV gain.

We now consider a scaled gain matrix based on Eq. (\ref{eqn:ssgain}). Let the \texttt{TC-REBOILER-SP} move size, $\Delta\textrm{MV}_1 = 2$, and the \texttt{FC-REFLUX-SP} move size, $\Delta\textrm{MV}_2 = 10$. Using typical move scaling, the gains in each CV now lie between -1 to +1. The scaled gain matrix, $G'$ is calculated using Eq. (\ref{eqn:svd_scaling}) as

\begin{equation}
\label{eqn:ssgain_scaled}
G' =
    \begin{bNiceMatrix}[first-row, last-col=3]
        {\footnotesize{\textrm{MV1}}} &  {\footnotesize{\textrm{MV2}}} & \\
        -1.0000 & -0.0754 & \footnotesize{\textrm{CV1}}  \\
        1.0000 & -0.7813 & \footnotesize{\textrm{CV2}}
    \end{bNiceMatrix}
\end{equation}

% \begin{equation}
% \label{eqn:ssgain_scaled}
% \renewcommand\arraystretch{1.5}
% \begin{blockarray}{rrr}
%                  & \rot{\texttt{TC-REBOILER}} &  \rot{\texttt{FC-REFLUX}} \\
%   \begin{block}{r[rr]}
%         \texttt{AI-BTMS-RVP}
%         & -1.0000 & -0.0754  \\
%         \texttt{AI-OVHD-C5} & 1.0000 & -0.7813 \\
%   \end{block}
% \end{blockarray}
% \end{equation}

% Figure \ref{fig:svd_scaled} shows the CV ellipse for the scaled gains. Clearly the 

Scaling has changed our understanding of the problem. $\gamma' = 2.68$ is now smaller for $G'$, compared to $\gamma = 11.7$ for $G$ prior to scaling. The higher the condition number, the harder it is to drive in the weak direction. From a distillation control perspective, we know that it is much easier to lower the RVP and increase the C5s simultaneously (shift the column composition profile u p), than it is to lower both the RVP and C5s (increase separation) \cite{sagforsMultivariableControlIllconditioned1998}.

\subsection{Relative Gain Array}
Inspecting the RGA value is another approach to identify gain ratio issues \cite{bristolNewMeasureInteraction1966}.  The advantage of RGA over SVD is that it is scale independent. In practice, for the gains in a multivariable controller matrix, the RGA is useful in setting an interaction threshold for significant and insignificant gain ratios. For the case of the $2\times2$ matrix, $G$ in Eq. (\ref{eqn:ssgain-general}), the RGA computations can be simplified \cite{brambillaMultivariableControllerDistillation1992} as:

% Through this analysis, an RGA above a certain threshold value would be removed. 
% RGA was originally developed to evaluate variable pairing in traditional advanced control pairing schemes .

\begin{equation}
\label{eqn:RGA}
\textsc{RGA} = \left(1 - \frac{G_{12}\cdot G_{21}}{G_{11}\cdot G_{22}}\right)^{-1}
\end{equation}

% This equation gives us a measurable degree of collinearity that is directly determined by the RGA; an infinite RGA tells us that the ratios are equal, and a zero RGA tells us that the equation is undefined and the CV rows are independent.

% Using the gains in $G$, the linear equations for the CV response based on MV moves can be rearranged as:
% \begin{equation}
% \label{eqn:CV1}
% \begin{aligned}
% CV_1 = G_{11} \cdot MV_1 + G_{12} \cdot MV_2 \\
% CV_2 = G_{21} \cdot MV_1 + G_{22} \cdot MV_2
% \end{aligned}
% \end{equation}

As the RGA number increases, the gain ratios in the $2\times2$ matrix approach each other and the matrix approaches singularity.  When the gains are collinear, the RGA number is infinite. For an APC model, it is useful to set an RGA threshold such that any gain ratios above a threshold RGA value need to be fixed.  For $G$ in Eq. (\ref{eqn:ssgain}), the RGA is calculated to be a desirable, small value of 1.11. In industrial practice, a reasonable 2x2 RGA threshold in a typical multivariable controller could be around 12. Because RGA is independent of scaling, the RGA describes process directionality with regard to inherent process stability, while SVD, because it is considering gain magnitude, does so with regard to performance \cite{sagforsMultivariableControlIllconditioned1998}. From an MPC standpoint, we may or may not be concerned from the performance point of view. Economically we still may want to use the MV with a small directional effect because it costs nothing to do so. For further details and control applications of RGA, the expository text by McAvoy \cite{mcavoyInteractionAnalysisPrinciples1983} is an excellent resource.

% Compared to SVD, RGA does not change with different engineering units or scaling methods which means that the RGA describes process directionality with regards to inherent process stability, while SVD does so with regards to performance \cite{sagforsMultivariableControlIllconditioned1998}. 

% Deshpande \cite{deshpandeMultivariableProcessControl1989} and 
% Note that it is possible to calculate a negative RGA number \cite{skogestadImplicationsLargeRGAelements1987}.  However, by simply changing the order of the equations, the calculated RGA number is always positive. 

\section{METHODOLOGY: BINNING THE GAINS}

% Moving beyond these toy examples, industrial gain matrices will typically have over 50 variables \cite{guerlain_mpc_2002}. As the size of the gain matrix increases, the potential for stability issues due to ill-conditioning increase exponentially because it applies to the gain matrix and all of its submatrices. In other words, a completely stable gain matrix $G$ with size $N\times N$ will have linearly independent submatrices of sizes $2 \times 2$, $3 \times 3, \ldots, N \times N$ in all permutations of $G$. Conditioning issues therefore become difficult to detect and tedious to mitigate as gain matrices become larger. 

%  such that all $2\times2$ submatrices in the gain matrix will always satisfy a predefined RGA.

We now present a novel, non-iterative methodology for gain matrix conditioning. The traditional approach involves tedious manual gain adjustments or a complicated optimization analysis that may not converge, until a reasonable RGA value is reached for all submatrices. Our methodology takes a user-defined RGA threshold to generate a discretized grid of $K$ gain values with desirable properties for matrix conditioning. %that can replace gains in ill-conditioned submatrices.

Given an $N\times M$ gain matrix, there are a total of ${M\choose 2} {\times} {N\choose 2}$ combinations of 2x2 submatrices. By expanding the factorials in the combination terms, it follows that the cost of calculating the RGA for all submatrices is $O(N^2 M^2)$. Our method mathematically guarantees that after a single-pass adjustment of all $N\times M$ gains, by matching each gain with the closest 'bin' in $K$ grid values, all 2x2 submatrices will have an RGA value below the threshold, with a time complexity of $O(NM \cdot \log K)$ using a binary search implementation, without the need to inspect individual submatrices.

% Binning also provides some benefits over the traditional conditioning approach; given $m$ MVs and $n$ CVs, checking all RGAs for conditioning issues is bounded by $O(m^2n^2)$ in both approaches. However, binning only requires a single pass of this calculation over all $2\times 2$ submatrices; traditional iteration may need many passes to ensure proper conditioning. Moreover, users can choose to forego this calculation and adjust all gains to the bin grid, bounding the complexity to $O(mnk)$ (with $k$ as the number of bin boundaries).

% The threshold used to define this grid ensures that all RGA values in the resulting $2 \times 2$ submatrices are smaller or equal.

% SHOW A CHART OF PERCENT CHANGE VS RGA?
% GOOGLE LATEX BUCKETS? HASH TABLE? BINNING? https://tex.stackexchange.com/questions/428893/errors-when-drawing-extendible-hash-index-table-with-tikz-library
% See fig in https://see.xidian.edu.cn/faculty/gmshi/paper/2012/wangTIP.pdf

% Consider a perfectly collinear gain matrix $G$ where one CV column is a multiple of another:

% \begin{equation}
% \label{eqn:collinear}
%     \frac{G_{11}}{G_{21}} = \frac{G_{12}}{G_{22}}
% \end{equation}

Consider any $2\times2$ non-zero gain matrix, $G$. The gains can be scaled to obtain $K$, such that 3 of the 4 gains are $1$, with a single non-unity gain, $k$, that is dependent on the matrix's RGA value:
\begin{equation}
\label{eqn:unityscaling}
\begin{aligned}
K &=
    \begin{bmatrix*}[c]
        1/G_{11} & 0 \\
        0 & 1/G_{21}
    \end{bmatrix*}
    \begin{bmatrix*}[c]
        G_{11} & G_{12} \\
        G_{21} & G_{22}
    \end{bmatrix*}
    \begin{bmatrix*}[c]
        1 & 0 \\
        0 & G_{21}/G_{22}
    \end{bmatrix*} \\
& = \begin{bmatrix*}[c]
        1 & k \\
        1 & 1
    \end{bmatrix*}, \quad \text{where } k = (1-\RGA^{-1})
\end{aligned}
\end{equation}

The $k$ term defines the relationship between the gain ratios and the RGA, as shown by rearranging Eq. (\ref{eqn:RGA}) to get

\begin{subequations}
\label{eqn:RGA2}
\begin{eqnarray}
G'_{12}/G'_{22} = (1 - \textsc{RGA}^{-1}) \cdot G'_{11}/G'_{21} \label{eqn:RGA2_col}\\
G'_{12}/G'_{11} = (1 - \textsc{RGA}^{-1}) \cdot G'_{22}/G'_{21} \label{eqn:RGA2_row}
\end{eqnarray}
\end{subequations}
where Eq. (\ref{eqn:RGA2_col}) defines the relationship for the column ratios, and likewise, Eq. (\ref{eqn:RGA2_row}) for the rows. For $\RGA = \infty$, $k=1$ and the matrix is singular, meaning that physically, either the MVs or CVs are collinear. For $\RGA = 12$, $k = 0.9167$. The intuition behind our method is that, instead of just using the matrix gains to compute $k$ and $\RGA$, we could also flip the relationship around to enforce a certain $\RGA$ value for any matrix. This is done by defining our desired $\RGA$, calculating $k$ and then adjusting the gains to equal $k$. 

% In an actual gain matrix, we have $k = \frac{G'_{12} G'_{21}}{G'_{22} G'_{11}}$, so $k$ is dependent on the physical situation represented by the gains. 

% In any typical-move scaled matrix obtained with Eq. (\ref{eqn:svd_scaling}), the absolute value of the scaled gains lie between $0$ to $1$. 
In any typical-move scaled matrix obtained with Eq. (\ref{eqn:svd_scaling}), the absolute values of the scaled gains lie between $0$ and $1$. Starting with the largest scaled gain of $1$, for $\RGA=12$, the next-largest value is $k=0.9167$ from Eq. (\ref{eqn:unityscaling}). To `condition' the matrix, gains between $1$ and $0.9167$ are adjusted to the closest value, either to $1$, enforcing collinearity, or to $0.9167$, enforcing $\RGA = 12$. 

To cover the entire range of typical move-scaled gains, we now seek a series of values, $k_1, \cdots, k_n$ such that all scaled gains can be adjusted to those values, and the entire matrix can be conditioned to any user-defined RGA threshold. We now ask, what is the next largest value of $k$ that will ensure that the matrix meets the desired $\RGA=12$ threshold? It is found by multiplying $K$ by $k$:

\begin{equation}
\label{eqn:unityscaling2}
\begin{aligned}
K^{(2)} &= k\cdot \begin{bmatrix*}[c]
        1 & k \\
        1 & 1
    \end{bmatrix*} 
= \begin{bmatrix*}[c]
        0.9167 & 0.8403 \\
        0.9167 & 0.9167
    \end{bmatrix*}    
\end{aligned}
\end{equation}

Similar to the first step, the absolute value of the scaled gains between $0.9167$ and $0.8403$ are adjusted to the closest value, either $0.9167$ or $0.8403$. We can repeat this procedure to get $K^{(3)}, \cdots, K^{(n)}$ until we generate enough values to cover the range of the scaled gain matrix.

Formally, the pattern in Eq. (\ref{eqn:unityscaling2}) can be used to construct a recurrence relation that yields a geometric sequence of gains that satisfy a desired RGA threshold. This sequence forms the basis of a set of discretized \textit{bins}. We define \textit{bins} as having strictly decreasing bin boundaries, $B_0, B_1, \cdots, B_n$ such that $B_{i+1} < B_{i}$ for $i = {0,1,\cdots,n}$. The \textit{bin boundaries} define the gain values we can use. The \textit{bin width} for each interval, $[B_{i+1}, B_{i}]$, is given by $w_i = B_{i} - B_{i+1}$. This definition constructs $n$ bins with $n+1$ bin boundaries. $n$ is chosen such that the bins cover the range of all gains in the matrix.

Let the first bin boundary be $B_{0} = 1$. The next term, $B_1$, is found with a recurrence relation derived from Eq. (\ref{eqn:unityscaling2}) as

$$B_1 = (1 - \textsc{RGA}^{-1}) \cdot B_{0}$$

% The procedure for generating the bin boundary values is defined as follows. Let the first bin boundary be $B_{0} = |G'_{11}/G'_{21}| = 1$, based on the scaled gain matrix in Eq. (\ref{eqn:svd_scaling}). The $i$\textsuperscript{th} bin boundary is calculated sequentially using a recurrence relation

% $$B_i = (1 - \textsc{RGA}^{-1}) \cdot B_{i-1}$$
Solving the recurrence relation yields a closed-form solution for any $i$\textsuperscript{th} bin boundary as a function of $B_0$ and RGA:

\begin{equation}
\label{eqn:RGA_buckets}
B_i = (1 - \textsc{RGA}^{-1})^i \cdot B_0
\end{equation}

% The $i$th bin interval is defined as the distance between its 2 neighboring boundaries,

% $$\Delta_i = B_i - B_{i-1}$$

% maybe follow this https://arxiv.org/pdf/2108.08228.pdf

Consider a scaled gain element $G'_{ij}$ in $G'$, with an absolute value $|G'_{ij}|$ that lies in the interval $B_{k+1} \leq |G'_{ij}| < B_{k}$. The binning procedure discretizes $G'_{ij}$ to obtain a \textit{binned gain}, $\widehat{G}_{ij}$, by assigning $G'_{ij}$ to the closest boundary with

% for all bins $k \in \{0,\cdots,n\}$

\begin{equation}
\label{eqn:binning_step}
\widehat{G}_{ij}= 
\sgn{(G'_{ij})} \times
\begin{cases}
    B_{k},& \text{if } |G'_{ij}| > \frac{B_k + B_{k+1}}{2}\\
    B_{k+1},              & \text{otherwise}
\end{cases}
\end{equation}

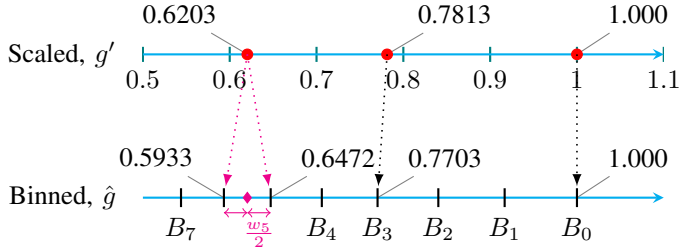
\begin{figure}
    \begin{tikzpicture}
        \begin{axis}[%
            name=first,
            clip=false,
            axis line style={thick, cyan},
            axis x line=middle,
            xticklabel style = {yshift=0cm},
            every x tick/.style={thick, color=teal},
            major tick length=0.20cm,
            axis y line=none,
            width=\axisdefaultwidth,
            xmin=0.5,
            xmax=1.1]
        \addplot [only marks,mark=*,red] coordinates {
            (0.7813,0) (1.0,0)
        };
        \addplot [only marks,mark=*,red] coordinates {
            (0.6203,0)
        };
        \coordinate[label=180:{Scaled, $g'$}](x0) at (axis cs:0.48,0);       
        \coordinate[pin={[pin edge={thin}, pin distance=0.4cm] above right:{1.000}}](x1) at (axis cs:1,0);
        \coordinate[pin={[pin edge={thin}, pin distance=0.4cm] above right:{0.7813}}](x2) at (axis cs:0.7813,0);
        \coordinate[pin={[pin edge={thin}, pin distance=0.4cm] above left:{0.6203}}](x3) at (axis cs:0.6203,0);        
        \end{axis} 
        
        \begin{scope}[yshift=-1.9cm] 
        \begin{axis}[%
            clip=false,
            ticks = none,
            axis line style={thick, cyan},
            axis x line=middle,
            major tick length=0.25cm,
            axis y line=none,
            xticklabel=\empty,
            width=\axisdefaultwidth,
            xmin=0.5,
            xmax=1.1]
        \addplot [only marks,mark=|,thick,black,mark size=0.125cm] coordinates {
            (0.5439,0) (0.5933,0) (0.6472,0) (0.7061,0) (0.7703,0) (0.8403,0) (0.9167,0) (1.0,0)
        };

        \addplot [only marks,mark=diamond*,magenta] coordinates {
            (0.6203,0)
        };
        
        \coordinate[label=180:{Binned, $\hat{g}$}](y0) at (axis cs:0.48,0);
        
        \def\dist{0.15cm}
        
        \coordinate[label={[label distance=\dist]270:{$B_7$}}](b7) at (axis cs:0.5439,0);
        % \coordinate[label={[label distance=\dist]270:{$B_6$}}](b7) at (axis cs:0.5933,0);
        % \coordinate[label={[label distance=\dist]270:{$B_5$}}](b7) at (axis cs:0.6472,0); 
        \coordinate[label={[label distance=\dist]270:{$B_4$}}](b7) at (axis cs:0.7061,0);
        \coordinate[label={[label distance=\dist]270:{$B_3$}}](b7) at (axis cs:0.7703,0);        
        \coordinate[label={[label distance=\dist]270:{$B_2$}}](b7) at (axis cs:0.8403,0);
        \coordinate[label={[label distance=\dist]270:{$B_1$}}](b7) at (axis cs:0.9167,0);
        \coordinate[label={[label distance=\dist]270:{$B_0$}}](b7) at (axis cs:1,0);
        
        \node (ymid) at (axis cs:0.6203,0) {};
        \node (ymidL) at (axis cs:0.5933,0) {};
        \node (ymidR) at (axis cs:0.6472,0) {};
        
        \coordinate[pin={[pin edge={thin}, pin distance=0.4cm] above right:{1.000}}](y1) at (axis cs:1,0);
        \coordinate[pin={[pin edge={thin}, pin distance=0.4cm] above right:{0.7703}}](y2) at (axis cs:0.7703,0);
        \coordinate[pin={[pin edge={thin}, pin distance=0.4cm] above left:{0.5933}}](y4) at (axis cs:0.5933,0);
        \coordinate[pin={[pin edge={thin}, pin distance=0.4cm] above right:{0.6472}}](y3) at (axis cs:0.6472,0);
        
        \end{axis}
        \end{scope}
        
        %numberline idea
        % https://www.google.com/search?q=number+line+braces+tikz&safe=active&rlz=1C1GCEU_enCA840CA840&sxsrf=AOaemvJxtPpFTjUdnEbJhMuB7VbNrKwtsg:1643214764847&source=lnms&tbm=isch&sa=X&ved=2ahUKEwjroeSC7M_1AhWWIDQIHfyzDIoQ_AUoAXoECAEQAw&biw=1494&bih=805&dpr=1.5#imgrc=Rl8x08SkqCYaDM&imgdii=kJF5qy5tyYMP1M
        
        \draw [-latex, dotted, shorten >=0.15cm, semithick] (x1) -- (y1);
        \draw [-latex, dotted, shorten >=0.15cm, semithick] (x2) -- (y2);
        \draw [-latex, dotted, magenta, shorten >=0.15cm, semithick] (x3) -- (y3);
        \draw [-latex, dotted, magenta, shorten >=0.15cm, semithick] (x3) -- (y4);
        
        % \draw [decorate, magenta, decoration={brace,amplitude=3.5pt,raise=0.2cm}]
        % (y3) -- (ymid) node [black,midway,yshift=-0.5cm] {\footnotesize
        % $\Delta$}; 
        
        % \draw [-, dashed, magenta, shorten >=-0.10cm, thick] (x3) -- (ymid);       
        % \draw [->, shorten >=0.1cm, thick, blue] (x3) to [out=0,in=5] (y3);
        
        \draw [to-to, magenta] ([yshift=-0.2cm]ymid.center) -- ([yshift=-0.2cm]ymidR.center) node[midway,below] {\footnotesize{$\frac{w_5}{2}$}};
        
        \draw [to-to, magenta] ([yshift=-0.2cm]ymid.center) -- ([yshift=-0.2cm]ymidL.center);        

    \end{tikzpicture}
    \caption{Illustration of the binning process. A gain, $g'$ that lies in the bin interval $[B_{i+1}, B_i]$ is adjusted to the closest bin boundary to output a binned gain, $\hat{g}$. Gains that lie on the midpoint of an interval are at equidistant from both bin boundaries, and will experience equal gain adjustment in either direction, as shown by the magenta arrows.}
    \label{fig:numberline}
\end{figure}

In other words, the procedure simply adjusts the typical-move scaled gain to the closest bin boundary, based on the absolute value of the gain. Due to the exponential term in Eq. (\ref{eqn:RGA_buckets}), the bin widths are also strictly decreasing, such that $w_{k+1} < w_{k}$. It follows that the maximum gain change introduced by this procedure is bounded by the width of the first bin interval $[B_1, B_0]$, with the largest gain adjustment occurring at the midpoint of $[B_1, B_0]$, denoted by $B_m$. As shown by the magenta marker in Figure \ref{fig:numberline}, a gain at the midpoint of $[B_6, B_5]$ could be assigned to either boundaries and receive an equal amount of gain adjustment.

% Note that this methodology constrains the change in gain ratio for a pair of gains, and constructs a discretization pattern for all gains in the matrix. Due to the scaling applied in Eq. (\ref{eqn:svd_scaling}), the maximum gain in the scaled matrix is 1. The astute reader may observe that some combinations of gains could result in gain ratios that are greater than 1 (e.g. by swapping $G'_{12}$ and $G'_{22}$ in Eqn. (\ref{eqn:ssgain_scaled})), but this does not affect the methodology since it is the gains themselves that are being adjusted to the nearest bins, not the ratios.

% Some combinations of gain values may result in gain ratios that are greater than 1 (e.g. if the gains for MV2 in Eqn. \ref{eqn:ssgain_scaled} were reversed), but due to the scaling applied in Eq. \ref{eqn:svd_scaling}, each CV row has a maximum gain of 1.

% These gain ratios still fit in this methodology, as the original scaling algorithm defines the maximum gains to have a magnitude of 1. The bins were constructed using this maximum gain, which means that we do not need to re-use the recurrence relation in Eqn. \ref{eqn:RGA_buckets} for each $2 \times 2$ submatrix.

% The discretized or binned gain matrix, $G'$ is given by applying the procedure on the absolute value of each gain element according to

% $$G'_{ij} = \sgn{(G_{ij})} \times Q(\left|G_{ij}\right|)$$
% Where $Q(x)$ is the binning function defined as the closest bin boundary $B_k$ to $x$.

The maximum relative gain change (\%), $\delta_{\max}$ is a function of only RGA and independent of the choice of $B_0$. By definition, the relative gain change is the ratio of the gain adjustment to the initial gain value, $B_m$

\begin{equation}
\label{eqn:relative_change}
    \delta_{\max} = \frac{B_0 - B_m}{B_m}\times 100\% \text{,} \quad B_m = \frac{B_0+B_1}{2}
\end{equation}

\begin{figure}
    \centering
    \begin{tikzpicture}[font=\footnotesize]
    \begin{axis}[
            width=\columnwidth,
            height=4.5cm,
            xmin=0, xmax=20, % x scale
            ymin=0, ymax=25, % y scale
            samples=20,
            minor y tick num=4,
            % ymajorgrids,
            % xmajorgrids,
            domain=1:20,  % added, key improvements
            xlabel={RGA choice},
            ylabel={$\delta_{\max}$ (\%)}
    ]
    \addplot {100*abs((1/x)/((1/x)-2))};
    \addplot [only marks,black,samples at={12}] {100*abs((1/x)/((1/x)-2))} node[pin=30:{$(12,4.35\%)$}]{};
    \addplot [only marks,black,samples at={6}] {100*abs((1/x)/((1/x)-2))} node[pin=30:{$(6,9.09\%)$}]{};
    \node[anchor=south west, text=black] (a) at (axis cs:8,17) {$\delta_{\max} = \dfrac{\RGA^{-1}}{2 -\RGA^{-1}} \times 100\%$};
    \end{axis}
    \end{tikzpicture}
    \caption{The maximum relative gain change increases exponentially with smaller RGAs, showing that larger gain adjustments are required to achieve lower RGA values. In practice, for a reasonable RGA threshold of 12, any gain change is guaranteed to at most 4.35\%.}
    \label{fig:changepercent}
\end{figure}
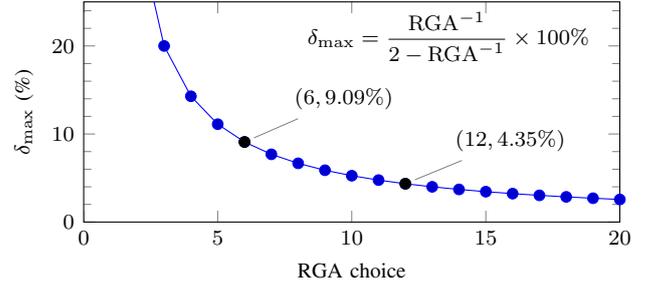

By substituting Eq. (\ref{eqn:RGA_buckets}) into Eq. (\ref{eqn:relative_change}), the $B_0$ terms cancel out and we get $
\delta_{\max} = \frac{\RGA^{-1}}{2 - \RGA^{-1}} \times 100\%$. The maximum relative gain change percentage as a function of RGA choice is shown in Figure \ref{fig:changepercent}. For an RGA choice of 12, the largest relative gain change percentage would have an upper bound of 4.34\%, where the 0.9583 midpoint gain is adjusted to either $B_1=0.9167$ or to $B_0 = 1.0$.

The selected RGA threshold builds a binning structure for gains to be fitted into. Note that applying the binning algorithm in a single pass over all gains will honor the RGA threshold for all 2x2 submatrices. Unlike the traditional method, we do not need to inspect individual submatrices and iteratively adjust the entire matrix to maintain the RGA threshold. If new gains are added to the gain matrix at a later time (e.g. controller revamps), they are simply adjusted to the existing binned structure. However, if certain gain values must be preserved (e.g. mass balance constraints), users can choose to adjust only the gains in submatrices that exceed the RGA threshold. All other gains can be left as-is.

% Users can also choose to adjust evaluate the RGA for each submatrix, if there's a need to preserve certain gain values such as mass balance constraints.

% It is also important to note that only the gains that cause $2 \times 2$ matrices to exceed the $\textsc{RGA}$ threshold need to be adjusted, and all others can be left as-is.

The binning procedure is applied to the scaled gain matrix, $G'$ in Eq. (\ref{eqn:ssgain_scaled}) to obtain binned gains $\widehat{G}$ as shown in Eq. (\ref{eqn:ssgain_binned}). CV1-MV2 was adjusted by 2.5\% and CV2-MV2 by 1.4\%.

\begin{equation}
\label{eqn:ssgain_binned}
\widehat{G} =
    \begin{bNiceMatrix}[first-row, last-col=3]
        {\footnotesize{\textrm{MV1}}} &  {\footnotesize{\textrm{MV2}}} & \\
        -1.0000 & -0.07351 & \footnotesize{\textrm{CV1}}  \\
        1.0000 & -0.77025 & \footnotesize{\textrm{CV2}}
    \end{bNiceMatrix}
\end{equation}

% \begin{equation}
% \label{eqn:ssgain_binned}
% \renewcommand\arraystretch{1.5}
% \begin{blockarray}{rrr}
%                  & \rot{\texttt{TC-REBOILER}} &  \rot{\texttt{FC-REFLUX}} \\
%   \begin{block}{r[rr]}
%         \texttt{AI-BTMS-RVP}
%         & -1.0000 & -0.07351  \\
%         \texttt{AI-OVHD-C5} & 1.0000 & -0.77025 \\
%   \end{block}
% \end{blockarray}
% \end{equation}

\section{CASE STUDY: DEBUTANIZER PROBLEM}

We apply these techniques on the debutanizer process described in Figure \ref{fig:debutanizer}. Although the operation of a debutanizer seems relatively simple, the process illustrates many complexities commonly faced by APC engineers. The raw gain matrix for the process is shown in Table \ref{tab:original} with the accompanying debutanizer schematic in Fig. \ref{fig:debutanizer}.

% The debutanizer example is based on simulated data originally developed at Controls Consulting Inc. (CCI) as part of an industrial APC class.

% The debutanizer example discussed here is based on simulated data originally developed by Dave Hoffman at Controls Consulting Inc. (CCI) about 20 years ago.  

\begin{table}[h]
\centering
\caption{\label{tab:original} Original Gains, $G$ }
% \textcolor{red}{\small{How do we have CVs for both PC-TOP-SP and PC-TOP-OP? }}
\resizebox{\columnwidth}{!}{%
    \begin{tabular}{l|rrrrr} 
    \toprule
    ~     & TC-REBOIL-SP & FC-REFLUX-SP & PC-TOP-SP & FC-DIST-SP & FI-FEED-PV  \\
    % ~      &      & \footnotesize{Reboiler SP} & \footnotesize{Reflux SP} & \footnotesize{Pressure SP} & \footnotesize{Dist. flow SP} & \footnotesize{Feed} \\
    \hline
    AI-RVP-PV~~~ &  $-0.1942$      & $-0.0029$      & $0.0711$       & $0$           & $0.0013$   \\
    AI-DIST-C5    &  $0.1843$       & $-0.0288$      & $-0.1907$      & $0$           & $0.0070$   \\
    TOP-PCT      &  $0.9220$       & $-0.1477$      & $-0.9458$      & $0$           & $0.0371$   \\
    LI-ACCUM-PV~ &  $0.2042$       & $0$            & $-0.0667$      & $-0.1485$     & $0.0381$   \\
    DP-DEBUT-PV~ &  $0.0774$       & $0.0063$       & $-0.0143$      & $0$           & $0.0064$   \\
    PC-TOP-OP~~ &  $4.9714$       & $0.5000$       & $-4.9887$      & $0$           & $0.3738$   \\
    FC-REBOIL-OP &  $4.5005$       & $0.3391$       & $-1.4486$      & $0$           & $0.2725$   \\
    FC-REFLUX-OP &  $0$            & $0.2651$       & $0$            & $0$           & $0$        \\
    \bottomrule
    \end{tabular}
}
\end{table}

    % AI-RVP-PV~~~ & \footnotesize{RVP} & $-0.1942$      & $-0.0029$      & $0.0711$       & $0$           & $0.0013$   \\
    % AI-DIST-C5    & \footnotesize{C5s} &  $0.1843$       & $-0.0288$      & $-0.1907$      & $0$           & $0.0070$   \\
    % TOP-PCT      & \footnotesize{Column P} &  $0.9220$       & $-0.1477$      & $-0.9458$      & $0$           & $0.0371$   \\
    % LI-ACCUM-PV~ & \footnotesize{Reflux Level} &  $0.2042$       & $0$            & $-0.0667$      & $-0.1485$     & $0.0381$   \\
    % DP-DEBUT-PV~ & \footnotesize{Column dP} &  $0.0774$       & $0.0063$       & $-0.0143$      & $0$           & $0.0064$   \\
    % PC-TOP-OP~~ & \footnotesize{Column P Output} &  $4.9714$       & $0.5000$       & $-4.9887$      & $0$           & $0.3738$   \\
    % FC-REBOIL-OP & \footnotesize{Reboiler Output} & $4.5005$       & $0.3391$       & $-1.4486$      & $0$           & $0.2725$   \\
    % FC-REFLUX-OP & \footnotesize{Reflux Output} & $0$            & $0.2651$       & $0$            & $0$           & $0$        \\

The raw gains are scaled according to their typical moves using Eq. (\ref{eqn:svd_scaling}), and analyzed using SVD and RGA to identify ill-conditioned $2\times2$ submatrices based on an RGA threshold of 12 and a corresponding condition number of 59, as shown in Table \ref{tab:scaled}. The markers indicate that the gain is part of an ill-conditioned $2\times 2$ submatrix based on SVD results (red stars, \markr), and also on RGA results (blue squares, \markb). As shown in Table \ref{tab:illconditioned}, there are 13 pairs above the condition number threshold, and 11 pairs above the RGA threshold. Notice that the first entry with a condition number of 59.1 has a corresponding RGA number of 14.4, while another pair with a condition number of 60 has an RGA number of 10.75. For higher-order sub-matrices, there are 34 $3 \times 3$ submatrices of gains with a condition number greater than 100, and 36 $4 \times 4$ submatrices with a condition number greater than 100. 

\begin{table}[h]
\centering
\caption{\label{tab:scaled} Scaled Gains, $G'$}
\resizebox{\columnwidth}{!}{%
    \begin{tabular}{l|rrrrr} 
    \toprule
    ~            & TC-REBOIL-SP & FC-REFLUX-SP & PC-TOP-SP & FC-DIST-SP & FI-FEED-PV  \\ 
    ~            & $\Delta\textrm{MV} = 2$ & $\Delta\textrm{MV} = 10$ & $\Delta\textrm{MV} = 2$ & $\Delta\textrm{MV} = 5$ & $\Delta\textrm{MV} = 10$ \\ 
    \hline
    AI-RVP-PV~~~ & $-1$     \pb\markr      &  $-0.0754$ \pb\pr         & $0.3664$  \pb\markr           & $0$   \pb\pr  & $0.0337$ \pb\pr            \\
    AI-DIST-C5    & $0.9666$ \markb \markr  &  $-0.7552$ \markb \markr  & $-1$ \markb \markr            & $0$   \pb\pr  & $0.1839$ \markb \markr     \\
    TOP-PCT      & $0.9748$ \markb \markr  &  $-0.7807$ \markb \markr  & $-1$ \markb \markr            & $0$   \pb\pr  & $0.1962$ \markb \markr     \\
    LI-ACCUM-PV~ & $0.5500$ \markb \markr  &  $0$       \pb\pr         & $-0.1797$ \markb \markr       & $-1$  \pb\pr  & $0.5129$ \pb\pr            \\
    DP-DEBUT-PV~ & $1$      \markb \markr  &  $0.4049$  \markb \markr  & $-0.1848$ \pb\pr              & $0$   \pb\pr  & $0.4145$ \pb\markr         \\
    PC-TOP-OP~~ & $0.9965$ \markb \markr  &  $0.5011$  \markb \markr  & $-1$ \markb \markr            & $0$   \pb\pr  & $0.3747$ \markb \markr     \\
    FC-REBOIL-OP & $1$      \markb \markr  &  $0.3767$  \markb \markr  & $-0.3219$ \markb \markr       & $0$   \pb\pr  & $0.3027$ \markb \markr     \\
    FC-REFLUX-OP & $0$      \pb\pr         &  $1$       \pb\pr         & $0$ \pb\pr                    & $0$   \pb\pr  & $0$      \pb\pr            \\
    \bottomrule
    \end{tabular}
}
\end{table}

% Table \ref{tab:illconditioned} shows the gains and $2\times 2$ pairs with condition numbers or RGA values above the selected threshold. There are 13 pairs above the condition number threshold, and 11 pairs above the RGA threshold. Notice that one condition number is 59.1 has an RGA number of 14.4, while another pair with a condition number of 60 has an RGA number of 10.75.  

\begin{table}
\centering
\caption{\label{tab:illconditioned} Ill-conditioned pairs in $G'$}
\resizebox{\columnwidth}{!}{%
\begin{tabular}{cccc|cc}
    \toprule
~  \textrm{MV1}  &   \textrm{MV2}    &    \textrm{CV1}   &  \textrm{CV2}  & $\gamma$ & RGA   \\
\hline
FC-REFLUX-SP & FI-FEED-PV~~ & PC-TOP-OP~~ & FC-REBOIL-OP & 59.14  & 14.36   \\
TC-REBOIL-SP & FI-FEED-PV~~ & DP-DEBUT-PV~ & PC-TOP-OP~~ & 59.99  & 10.75   \\
TC-REBOIL-SP & PC-TOP-SP~~~ & AI-RVP-PV~~~ & LI-ACCUM-PV~ & 67.50  & 9.26    \\
TC-REBOIL-SP & FC-REFLUX-SP & DP-DEBUT-PV~ & FC-REBOIL-OP & 81.83  & 14.37   \\
FC-REFLUX-SP & PC-TOP-SP~~~ & AI-DIST-C5    & TOP-PCT~~    & 124.38 & 30.54   \\
TC-REBOIL-SP & PC-TOP-SP~~~ & AI-DIST-C5    & PC-TOP-OP~~ & 131.01 & 33.24   \\
TC-REBOIL-SP & FC-REFLUX-SP & AI-DIST-C5    & TOP-PCT~~    & 165.64 & 40.79   \\
PC-TOP-SP~~~ & FI-FEED-PV~~ & AI-DIST-C5    & TOP-PCT~~    & 169.40 & 16.04   \\
TC-REBOIL-SP & PC-TOP-SP~~~ & TOP-PCT~~    & PC-TOP-OP~~ & 181.27 & 45.81   \\
TC-REBOIL-SP & FI-FEED-PV~~ & AI-DIST-C5    & TOP-PCT~~    & 189.76 & 18.39   \\
FC-REFLUX-SP & FI-FEED-PV~~ & AI-DIST-C5    & TOP-PCT~~    & 276.03 & 32.66   \\
TC-REBOIL-SP & PC-TOP-SP~~~ & AI-DIST-C5    & TOP-PCT~~    & 472.37 & 118.54  \\
TC-REBOIL-SP & PC-TOP-SP~~~ & LI-ACCUM-PV~ & FC-REBOIL-OP & 530.00 & 66.23   \\
\bottomrule
\end{tabular}
}
\end{table}

% The FI-FEED-PV was originally a feedforward; here it is left in the gain matrix to increase the model complexity.  

Applying the binning strategy of Eq. (\ref{eqn:binning_step}) to the scaled gains yields the binned gains in Table \ref{tab:binned}, which shows each gain's percentage change in red. The resulting gain matrix has three $2 \times 2$ submatrices with an SVD condition number above 59. To eliminate the pairs identified by the condition number threshold, larger gain changes would be required. One of the SVD-identified pairs has an RGA of 8.4. It is therefore unreasonable to remove or adjust the gains associated with this pair because of a somewhat arbitrary SVD condition number. Waller and Waller \cite{wallerDefiningDirectionalityUse1995a} came to the same conclusion related to using the condition number as a measure of ill-conditioning decades ago.

% We now apply the binning strategy and show the binned gains in Table \ref{tab:binned} with the \% change in each gain in red. If we use SVD to analyze the 2x2s, there are 3 pairs of gains with condition numbers above 59.  To clear those condition numbers, the RGAs for those gains would have to be reduced below 12. As the scaling on the MVs and CVs is changed, the gain pairs above/below a condition number threshold also change.  Because of the scale independent nature of the RGA calculation, the threshold seems more characteristic of the degree of interaction between the gains.  Historically we have spent a lot of time trying to come up with the “proper” scaling which would allow SVD to solve the problem. Using RGA, the scaling issue is eliminated from the problem.

\begin{table}
\centering
\caption{\label{tab:binned} Binned Gains, $\widehat{G}$}
\resizebox{\columnwidth}{!}{%
\begin{tabular}{l|rrrrr}
    \toprule
    ~            & TC-REBOIL-SP & FC-REFLUX-SP & PC-TOP-SP & FC-DIST-SP & FI-FEED-PV  \\
    \hline
    AI-RVP-PV~~~ & $-1$                   & $-0.0754$                   & $0.3664$                & $0$       & $0.0337$            \\
    AI-DIST-C5    & $1$ \rt{3.46}          & $-0.7703$ \rt{2.00}         & $-1$                    & $0$       & $0.1914$ \rt{4.08}  \\
    TOP-PCT~~    & $1$ \rt{2.59}          & $-0.7703$ \rt{-1.34}        & $-1$                    & $0$       & $0.1914$ \rt{-2.41} \\
    LI-ACCUM-PV~ & $0.5439$ \rt{-1.11}    & $0$                         & $-0.1755$ \rt{-2.37}    & $-1$      & $0.5129$            \\
    DP-DEBUT-PV~ & $1$                    & $0.4189$  \rt{3.46}         & $-0.1848$               & $0$       & $0.4189$ \rt{1.06}  \\
    PC-TOP-OP~~ & $1$ \rt{0.35}          & $0.4985$  \rt{-0.52}        & $-1$                    & $0$       & $0.3840$ \rt{2.49}  \\
    FC-REBOIL-OP & $1$                    & $0.3840$  \rt{1.93}         & $-0.3227$ \rt{0.25}     & $0$       & $0.2958$ \rt{-2.30} \\
    FC-REFLUX-OP & $0$                    & $1$                         & $0$                     & $0$       & $0$                 \\
    \bottomrule
\end{tabular}
}
\end{table}

Reviewing the matrix, the binned gain matrix has 10 collinear $2 \times 2$ pairs, where there were none in the raw matrix. Table \ref{tab:collinear} highlights the new collinear pairings.

\begin{table}
\centering
\caption{\label{tab:collinear} Collinear Pairs in $\widehat{G}$}
\resizebox{\columnwidth}{!}{%
    \begin{tabular}{c|cccc}
        \toprule
        Pair & \textrm{MV1}  &   \textrm{MV2}    &    \textrm{CV1}   &  \textrm{CV2}  \\
        \hline
        1 & TC-REBOIL-SP & FC-REFLUX-SP & AI-DIST-C5 & TOP-PCT~~  \\
        2 & TC-REBOIL-SP & PC-TOP-SP~~~ & AI-DIST-C5 & TOP-PCT~~  \\
        3 & TC-REBOIL-SP & FI-FEED-PV~~ & AI-DIST-C5 & TOP-PCT~~  \\
        4 & FC-REFLUX-SP & PC-TOP-SP~~~ & AI-DIST-C5 & TOP-PCT~~  \\
        5 & FC-REFLUX-SP & FI-FEED-PV~~ & AI-DIST-C5 & TOP-PCT~~  \\
        6 & PC-TOP-SP~~~ & FI-FEED-PV~~ & AI-DIST-C5 & TOP-PCT~~  \\
        7 & FC-REFLUX-SP & FI-FEED-PV~~ & PC-TOP-OP~~ & FC-REBOIL-OP  \\
        8 & TC-REBOIL-SP & PC-TOP-SP~~~ & LI-ACCUM-PV~ & FC-REBOIL-OP  \\
        9 & TC-REBOIL-SP & PC-TOP-SP~~~ & AI-DIST-C5 & PC-TOP-OP~~  \\
        10 & TC-REBOIL-SP & PC-TOP-SP~~~ & TOP-PCT~~ & PC-TOP-OP~~  \\
        \bottomrule
    \end{tabular}
}
\end{table}

Notably, the number of $3\times 3$ gain combinations in the binned gain matrix above a condition number of 100 has been reduced from 34 in the raw matrix to two. The condition numbers of the two $3 \times 3$ matrices are 135 and 105. Similarly, the number of $4 \times 4$ gain combinations with a higher condition number than 100 has been reduced from 36 to 1, where that condition number is 156. By conditioning the $2 \times 2$ gains, the conditioning of the entire gain matrix has improved significantly. It is important to note that a high condition number on a higher-order matrix does not necessarily mean that this combination of gains will be active in an LP solution. Still, reducing the high condition number interactions means that it is impossible for such combinations to play a part in the LP solution under any situation.

% after binning the 2x2 gains is now 2 (versus 34 for the original gains). The condition numbers for the 2 3x3 combinations are 105 and 135. The number of 4x4 gain combinations above a condition number of 100 is 1 (versus 36 for the original gains) and its condition number is 156.   By conditioning the 2x2 gains, the conditioning of the entire gain matrix has been improved significantly. Reviewing the gain matrix statistics, there are 105 2x2 pairs of gains in the model.  For the 3x3 analysis, there are 311 sets of 3x3 gains.  Many of the 3x3 gain combinations are not likely to be part of an LP solution for the controller.

\section{DISCUSSIONS}

% \begin{itemize}
%     \item identifying combinations that \textit{should} be collinear from a first-principles point of view
%     \item SVD visualization/difficulty to interpret for larger matrices
% \end{itemize}

How do we know if a gain pair should be collinear or not collinear? Often the answer relies on process knowledge. Second to process understanding, if a gain pair is nearly collinear, removing the degree of freedom based on a high RGA or condition number is reasonable.

In the debutanizer example, the \texttt{TOP-PCT} is a pressure compensated temperature, and it should be collinear with the \texttt{AI-DIST-C5} composition. In the binned matrix, it is collinear since the scaled gains are equal. However, if one or more of the gains had been slightly bigger or smaller, the binning process would put the gain in a bin above or below the correct bin. If this were to happen, a degree of freedom would exist in the model where none should be. The opposite situation is also possible, where a degree of freedom is removed in a set of gains where one should exist.

% Improving the conditioning of a set of gains is also an option.

Under certain circumstances, ill-conditioned gains may not be a problem in the controller. Why would we even include the \texttt{TOP-PCT} in the model if we have \texttt{AI-DIST-C5} which serves the same purpose of measuring C5 composition? One reason is the C5 analyzer may not be reliable. If the controller is intermittently hitting the \texttt{AI-DIST-C5} or \texttt{TOP-PCT} upper limit, with an unconditioned gain matrix, the LP may believe that temporarily it can control both CVs. As long as the CV limits are spread widely, and the gains are nearly collinear, there likely will be no observed problems. The RGA number is high, but the LP is not trading off the degrees of freedom that exist between the two CVs.

% There are other options for dealing with this problem, but this is one option.

Conditioning problems for this case become apparent when the analyzer starts to drift, and the controller tries to control one CV at a high limit, and the other at a low limit.  A properly conditioned model will have no difference in the gain ratios of the two CVs, and when the LP is faced with conflicting limits, it will simply give up on the least important CV.  An unconditioned model will find small degrees of freedom that will allow the LP to trade off MV moves that try to fix the problem.  If the CV gain ratios are a little farther from collinearity, the moves to control CVs at different limits may not be large, but the LP will drive the process in unreasonable directions and CV errors will not be reduced. When differences in gain ratios of collinear variables exist, the controller LP may find a different solution based on which CV limit is active.  This can lead to optimization cost tuning in order to get the controller to behave the same way when faced with high or low limits on either variable.

The gain issues described can spread to higher dimensions as well, and the issues are often magnified when the optimizer becomes infeasible.  As long as the MV directions that try to solve the infeasibility are the same as the feasible economic direction, little changes may be noticed when the solution becomes infeasible.  When the infeasible direction opposes the economic direction, the controller can make large moves in an uneconomic direction trying to reduce or eliminate CV error.  From these behaviors, the practitioner may erroneously conclude that the model is inaccurate and needs re-evaluation.  A ``good" but unconditioned model can account for this behavior. Retesting to fix an ill-conditioned matrix is unlikely to solve the problem.

These simple examples show why binning the gains is not the final answer, rather the first step in the overall model gain analysis. The ``correct" structure of the gain matrix must be determined based on plant data and process understanding. The binning procedure makes the next step of validating the model gain structure much easier. In a real-world MPC project, the practitioner must perform an interaction analysis of the model gains by simulating the LP solution. 

\section{CONCLUSIONS}

In this paper, we have demonstrated a binning procedure to condition the gain matrix used in industrial MPCs. By conditioning the matrix, we remove small degrees of freedom that are either insignificant or zero, without large perturbations to the process gains. This gain conditioning step prevents the optimizer from erroneously exploiting small degrees of freedom for control. The binning approach can also include small gains in the model to maintain gain consistency and respect material balances, without introducing further conditioning problems.

There are several areas of interesting future work here. The tables presented earlier for identifying gain interactions using markers are quite basic - as these tables grow in size, this colour-based strategy can result in cluttered and unintuitive displays, making near-collinearity analysis visually difficult. There is an opportunity for improving the visualization of degrees of freedom and gain ratios. We have made some recent progress in visualizing the LP solution for troubleshooting and understanding online MPCs \cite{elnawawiInteractiveVisualizationDiagnosis2022}.

% There are several areas of interesting future work here.  The LP is driven by the gain ratios in a model. The tables presented earlier for identifying gain interactions with red and blue markers are very crude. In industry, we have seen practitioners build spreadsheets that use colors to identify different gain ratios. For larger matrices, this color-based strategy breaks down and produces a cluttered display, making it difficult to analyze the near-collinear pairs. There is an opportunity for improvement in visualization of the model's degrees of freedom and gain ratios. We've made some recent progress in visualizing the LP solution for troubleshooting online MPCs and understanding controller actions .

% Next, although 2x2 gain interactions can be solved with binning, it is possible there are model gain issues that seem to require setting the value of a gain in between bins. Depending on the actual model structure, making the change may or may not introduce a gain conditioning problem in the model.  If it does, it can be a difficult problem to solve.

Unlike SVD, higher-order RGA calculations rely on the matrix inverse, which does not exist for a rank-deficient matrix with collinearities. Therefore, to identify conditioning issues in larger submatrices, the RGA is mathematically undefined and cannot be used, and SVD may not be entirely reliable either due to scaling issues. In industry, higher-dimensional analysis of the gain matrix beyond $3\times3$ submatrices are often not investigated by practitioners. It may be that a higher order RGA analysis could be useful once all $2\times2$ submatrices have been conditioned to the desired RGA threshold. The goal, in practice, has always been to certify that a model is conditioned to the dimensionality of the problem.  Here the dimensionality is defined as a square case, corresponding to the lesser of the number of MVs or the number CVs. For a non-square SVD case, the extra variables generally guarantee that there are significant directions related to the degrees of freedom in the matrix. With SVD, it seems reasonable that for a specific problem, we should be able to identify physically meaningful orthogonal directions in the model, and then force the model gains to follow combinations of these directions.

% Lastly, we speculate that by using SVD with the MVs scaled by the optimization costs, we should be able to pair MVs and CVs, something users of the technology have always desired. We have come close to good results with this approach, but there are still cases when it seems to fail, and warrants further research.

\addtolength{\textheight}{-12cm}  % This command serves to balance the column lengths
                                  % on the last page of the document manually. It shortens
                                  % the textheight of the last page by a suitable amount.
                                  % This command does not take effect until the next page
                                  % so it should come on the page before the last. Make
                                  % sure that you do not shorten the textheight too much.

\bibliographystyle{IEEEtran}
\bibliography{IEEEabrv,references}

\end{document}